\documentclass[11pt,a4paper]{article}
\usepackage{amsmath}

\usepackage[psamsfonts]{amssymb}
\usepackage{amsmath}
\usepackage{epsfig}

\author{H. Mohseni Sadjadi\footnote{mohsenisad@ut.ac.ir}
\\ {\small Department of Physics, University of Tehran,}
\\ {\small P. O. B. 14395-547, Tehran 14399-55961, Iran}}
\title{ $w_d=-1$ in interacting quintessence model }

\begin{document}
\maketitle
\begin{abstract}
A model consisting of quintessence scalar field interacting with
cold dark matter is considered. Conditions required to reach 
$w_d=-1$ are discussed. It is  shown that depending on the
potential considered for the quintessence, reaching the phantom
divide line puts some constraints on the interaction between dark
energy and dark matter. This also may determine the ratio of dark
matter to dark energy density at $w_d=-1$.

\end{abstract}

\section{Introduction}

Dark energy model is a candidate to explain the present
accelerated expansion of the universe \cite{acc}. In this scenario
70\% of the universe is assumed to be permeated by a smooth energy
component with negative pressure, dubbed as dark energy. A simple
model introduced to describe dark energy is a scalar dynamical
field with a suitable potential \cite{scalar}.  In this model an
exact solution for Friedmann equations is only accessible for some
kinds of potentials \cite{bar}.

Based on recent astrophysical data, it seems that the dark energy
component has an equation of state parameter, $w_d\simeq -1$, in
the present epoch \cite{mone}. Therefore to study the dynamical
behavior of dark energy in the present era, instead of trying to
find an exact solution, one can restrict himself to the region
$w_d\simeq -1$, where equations can be solved approximately.

In a noninteracting dark energy model, it is expected that density
of dark matter ($\rho_m$), in the present epoch, be very less than
the density of dark energy component ($\rho_d$). This lies on the
fact that the equation of state (EoS) parameter of dark energy,
$w_d$, is less than $-{1\over 3}$, therefore it redshifts more
slowly than the (dark) matter component. But the ratio of dark
matter to dark energy, $r$, is of order unity in the present
epoch: $r\sim 3/7$, this is known as the coincidence problem
\cite{coin}. This forces us to consider interaction between dark
energy and matter which allows energy exchange between these
components \cite{inter}.

In this paper we assume that the universe is filled with a scalar
field, whose EoS satisfies $-1\leq w_d< -{1\over 3}$ (dubbed as
quintessence), and (cold) dark matter with mutual
non-gravitational interaction. As we have mentioned obtaining an
exact solution for the Friedman equations in the presence of
interactions even for simple potentials is not straightforward.
Hence using some natural conditions which must be obeyed by the
quintessence field in the vicinity of $w_d\simeq -1$, we seek the
required conditions (posed on the interactions and scalar
potentials) allowing  the system to reach at $w_d=-1$ without
requirement to obtain an analytic and an exact solution for the
problem. So using this method one can show that if a proposed
quintessence model with a specific potential and interaction
permits reaching to $w_d\simeq -1$ in the present era or not.

Besides, conditions  on the interaction term and the potential
which may be functions of energy densities may pose some natural
conditions on the ratio of dark matter to dark energy density
showing whether the occurrence of coincidence problem near the era
where $w_d=-1$ is occurred is only a coincidence.

 At the end, using the interaction $Q=H(\lambda_m
\rho_m+\lambda_d\rho_d)$ and via Taylor series expansion we
illustrate and emphasize our results.

We use units $G=k_{B}=c=1$ throughout the paper.

\section{Quintessence model and $w_d=-1$}
The spatially flat Friedmann Robertson walker space time is
described by the metric
\begin{equation}\label{1}
ds^2=dt^2-a^2(t)(dx^2+dy^2+dz^2),
\end{equation}
where $a(t)$ is the scale factor.  We assume that this universe is
filled by dark energy and (cold) dark matter with densities
$\rho_d$ and $\rho_m$ respectively. Dark energy component is a
scalar field, $\phi$, with potential $V( \phi)$.  Energy density,
$\rho_d>0$, and pressure, $P_d<0$,  of dark energy are given by
\begin{eqnarray}\label{2}
\rho_d&=&{{\dot{\phi}}^2\over 2}+V(\phi)\nonumber\\
P_d&=&{{\dot{\phi}^2}\over 2}-V(\phi).
\end{eqnarray}
The EoS parameter of dark energy, given by
\begin{equation}\label{3}
w_d={P_d\over \rho_d}={{\dot{\phi}}^2-2V(\phi)\over
{\dot{\phi}}^2+2V(\phi)},
\end{equation}
satisfies $-1\leq w_d< -{1\over 3}$. The scalar field with this
EoS parameter is dubbed quintessence. The inequality $w_d<-{1\over
3}$ is necessary for accelerating the expansion of the universe.
The Hubble parameter, $H={\dot{a}\over a}$, satisfies Friedmann
equations
\begin{eqnarray}\label{4}
H^2={8\pi\over 3}\rho\nonumber \\
\dot{H}=-4\pi(P+\rho),
\end{eqnarray}
where $\rho$ and $P$ are the total energy density and pressure of
the universe: $\rho=\rho_d+\rho_m$, $P=P_d$. We consider an
interaction between dark matter and dark energy components:
\begin{eqnarray}\label{5}
\dot{\rho_d}+3H(P_d+\rho_d)&=&-Q\nonumber \\
\dot{\rho_m}+3H\rho_m&=&Q.
\end{eqnarray}
By substituting (\ref{2}) in the above equation we find the
evolution equation for the quintessence field
\begin{eqnarray}\label{6}
\dot{\phi}(\ddot{\phi}+3H\dot{\phi}+V'(\phi))=-Q.
\end{eqnarray}
The ratio of dark matter to dark energy, $r={\rho_m\over \rho_d}$,
satisfies
\begin{equation}\label{7}
\dot{r}=3Hrw_d+{Q\over {\rho_d}}(1+r).
\end{equation}
Using $w_d<-{1\over 3}$ and the above equation one can show that
in noninteracting quintessence model: $\dot{r}<-Hr$, and therefore
$r\to 0$ eventually, in contrast to the recent data which assess
$r\sim \mathcal{O}(1)$. We can also obtain an expression for time
evolution of $r$ in terms of $\Omega_d$ defined by
$\Omega_d={\rho_d\over \rho_c}$ (where $\rho_c={3H^2\over 8\pi}$
is the critical density):
\begin{equation}\label{8}
\dot{r}=-{\dot{\Omega_d}\over \Omega_d^2}.
\end{equation}
By comparing (\ref{7}) and (\ref{8}) we arrive at
\begin{equation}\label{9}
w_d=-{\dot{\Omega_d}\over 3H\Omega_d(1-\Omega_d)}-{Q\over
3H\rho_d(1-\Omega_d)}.
\end{equation}
In the absence of interaction we have
$\dot{\Omega_d}>H(1-\Omega_d)$, which implies that $\Omega_d$ is
an increasing function of time. In contrast, $\Omega_m$, defined
by $\Omega_m={\rho_m\over \rho_c}$, is a decreasing function in
this case. Hence the introduction of $Q$ in the above equation,
via exchanging energy, depending on the form of $Q$ may prevent
$r={\Omega_m\over \Omega_d}$ to go to zero, and can be a remedy
for the coincidence problem.

Eq. (\ref{9}) is a general equation for dark energy models. In the
scalar field quintessence model we can also make use of
\begin{eqnarray}\label{10}
2V(\phi)&=&(1-w_d)\rho_d \nonumber \\
\dot{\phi}^2&=&(1+w_d)\rho_d,
\end{eqnarray}
to obtain
\begin{equation}\label{11}
[(1-w_d)\rho_d \dot{]}=2V'(\phi)\dot{\phi},
\end{equation}
and subsequently
\begin{equation}\label{12}
\pm\sqrt{(1+w_d)\rho_d}V'(\phi)={1\over
2}(-\dot{w_d}\rho_d+(1-w_d)\dot{\rho_d}).
\end{equation}
$+(-)$ corresponds to $\dot{\phi}\geq(\leq)0$. By substituting
\begin{equation}\label{13}
\dot{\rho_d}={3H^2\over 8\pi}\left(
-3H\Omega_d(1+w_d\Omega_d)+\dot{\Omega_d}\right),
\end{equation}
which can be verified by taking the time derivative of
$\rho_d=\Omega_d\rho_c$, into (\ref{12}), and by making use of
(\ref{9}), we arrive at
\begin{equation}\label{14}
\dot{w_d}=\mp 2V'(\phi)\sqrt{{1+w_d\over
\rho_d}}-3H(1-w_d^2)-{Q\over \rho_d}(1-w_d).
\end{equation}
We use the above equation to study the behavior of the model in
the vicinity of the time, $t=t_0$,  when $w_d(t_0)=-1$. The
equation of state parameter of the quintessence is equal to or
greater than $-1$, $w_d\geq -1$, therefore at $t_0$ we must have
$\dot{w_d}(t_0)=0$. Otherwise there will be a neighborhood of
$t_0$, in which $w_d<-1$. Hence a necessary condition to reach at
$w_d(t_0)=-1$ in the quintessence model is $\dot{w_d}=0$ at
$t=t_0$. Even when $t_0\to \infty$, this is asymptotically valid.
This lies on the fact if $w_d\to -1$ when $t\to \infty$, we must
also have ${dw_d\over dt}(\in \Re)=0$. For $\lim_{w_d\to
-1}{V'(\phi)\over \sqrt{V(\phi)}}\sqrt{(1+w_d)}=0$ (e.g. for
bounded ${V'(\phi)\over \sqrt{V(\phi)}}$ at $t=t_0$) (\ref{14})
reduces to
\begin{equation}\label{16}
{Q(t_0)\over \rho_d(t_0)}=0.
\end{equation}
E.g. if one takes the interaction term such as $Q=\lambda
\rho_m\rho_d$ $(\,\,\, \lambda>0)$, as in the present epoch $r\sim
\mathcal{O}(1)$, we have ${Q\over \rho_d}\neq 0$ and it is clear
that $w_d=-1$ cannot occur in the present era.

Note that because of the presence of additional terms in
(\ref{6})with probable singular behavior at the limit $w_d \to
-1$, $Q(t_0)=0$ (at $w_d=-1$) may not be derived directly from
(\ref{6}) unless $\lim_{w_d\to -1}{V'(\phi)\over
\sqrt{V(\phi)}}\sqrt{(1+w_d)}=0$.

Beside the above natural condition, $\ddot{w_d}(t_0)\geq 0$ is
also a necessary condition to reach at $w_d(t_0)=-1$. This is due
to the fact that $w_d=-1$ may only be the global minimum of
$w_d(t)$. Note that even if $w_d=-1$ occurs at $t_0\to \infty$,
$lim_{t\to \infty}\ddot{w_d}(t)=0$ must be satisfied (provided
$\ddot{w_d}\in \Re$). Now we must explain this natural condition
in terms of the interaction term and the potential (inconsistency
of the potential and the considered interaction with this
condition indicates that the model is not able to reach at
$w_d=-1$). To do so by getting another time derivative of
eq.(\ref{14}), at $t=t_0$ we obtain
\begin{equation}\label{17}
\ddot{w_d}(t_0)=-2{d\left({Q\over \rho_d}\right)\over {dt}} \mp
2{d\left({V'(\phi)\over \sqrt{\rho_d}}({\sqrt{1+w_d}})\right)\over
{dt}}.
\end{equation}
Note that the right hand side must be evaluated at $t=t_0$.  By
putting (\ref{14}) into (\ref{17}) we get
\begin{equation}\label{18}
\ddot{w_d}(t_0)=-2{d\left({Q\over \rho_d}\right)\over
{dt}}+2{(V'(\phi))^2\over V(\phi)} \pm 2{V'(\phi)\over
\sqrt{V(\phi)}}{Q\over \rho_d\sqrt{1+w_d}}.
\end{equation}
But at $t=t_0$
\begin{eqnarray}\label{19}
{d\left({Q\over \rho_d}\right)\over {dt}}&=&{\dot{Q}\over \rho_d}+\left({Q\over \rho_d}\right)^2\nonumber \\
&=&{\dot{Q}\over \rho_d}
\end{eqnarray}
holds, therefore $\ddot{w_d}(t_0)>0$ requires that at $t=t_0$:
\begin{equation}\label{20}
-\dot{Q}(t_0)+(V'(\phi))^2\pm {V'(\phi)\over
\sqrt{V(\phi)}}{Q(t_0)\over \sqrt{1+w_d}}\geq 0.
\end{equation}
For interactions of the form $Q(\rho_m,\rho_d)$, at $t=t_0$ we
have $\dot{Q}(t_0)=-3H\rho_m{\partial Q\over \partial \rho_m}$ and
the equation (\ref{20}) at $t=t_0$ reduces to
\begin{equation}
-3H\rho_m{\partial Q\over \partial \rho_m} +(V'(\phi))^2\pm
{V'(\phi)\over \sqrt{V(\phi)}}{Q\over \sqrt{1+w_d}}\geq 0.
\end{equation}

Note that the form used in (\ref{18}) for the third term in the
right hand side is suitable only when $Q$ can be expressed in
terms of $1+w_d$, (e.g. for interactions containing positive power
of $\dot{\phi}$ which following (\ref{10}) can be written in terms
of $1+w_d$ like the interaction considered for the inflaton during
the reheating process :$Q\propto P_d+\rho_d=\dot{\phi}^2$). In
general, at $t=t_0$, using l'H\^{o}pital's rule, it is also
possible to write (\ref{18}) in the form
\begin{equation}\label{21}
\ddot{w_d}(t_0)=-2{d\left({Q\over \rho_d}\right)\over
{dt}}\mp{V'(\phi)\over \sqrt{V(\phi)}}\sqrt{2\ddot{w_d}(t_0)}.
\end{equation}
For intermediate time (we mean $t_0\nrightarrow \infty$), and for
$t<t_0$, we have $\dot{w_d}<0$ and for $t>t_0$, $\dot{w_d}>0$
holds. $\{+ (-)\}$ corresponds to the case where $\dot{\phi}\leq
(\geq)0$ when $t\leq t_0$. (\ref{21}) has real roots (for
$\ddot{w_d}$) provided that at $t=t_0$
\begin{equation}\label{22}
4\dot{Q}(t_0)\leq \left(V'(\phi)\right)^2.
\end{equation}
The above inequality can be viewed as a constraint on the
parameters of the model. In general, if the first nonzero
derivative of $w_d$ at $t_0$ is of order $n$, at $t=t_0$ we have
\begin{equation}\label{23}
w_d^{(n)}(t_0)=-2\left({Q\over \rho_d}\right)^{(n-1)}\mp
2\left({V'(\phi)\over
\sqrt{\rho_d}}({\sqrt{1+w_d(t_0)}})\right)^{(n-1)}.
\end{equation}
In this case, evenness of $n$ and $w_d^{(n)}(t_0)>0$, together
with $\dot{w_d}(t_0)=0$ are sufficient conditions for $w_d$ to
have a global minimum at $t_0$. The generalization of (\ref{21})
is then
\begin{equation}\label{24}
w_d^{(n)}(t_0)=-2\left(Q\over \rho_d\right)^{(n-1)}\mp
2{(n-1)!\over ({n\over 2}-1)!}\left({V'(\phi)\over
\sqrt{\rho_d}}\right)^{({n\over 2}-1)}\sqrt{{w_d^{(n)}(t_0)\over
n!}}.
\end{equation}
To derive the above equation we have assumed that
$\left({V'(\phi)\over \sqrt{\rho_d}}\right)$ and its time
derivatives up to order $({n\over 2}-1)$ are continuous and
bounded at $t=t_0$. If $w_d$ tends asymptotically to $-1$, all of
derivatives of $w_d$ may be zero in this limit. In this situation
$t_0$ is the point of infinite flatness and for infinitely
differentiable $w_d$, can only occur at infinity, $t_0\to \infty$.

In the following, to elucidate our results, as an example, we
consider the interaction \cite{examp}
\begin{equation}\label{25}
Q=H(\lambda_m\rho_m+\lambda_d\rho_d).
\end{equation}
Following (\ref{16}), we deduce that in order to reach at
$w_d=-1$, we must have
\begin{equation}\label{26}
r(t_0)=-{\lambda_d\over \lambda_m}.
\end{equation}
This equation determines the ratio of dark matter to dark energy
in terms of the parameter of interaction at $t=t_0$ . As we
consider $r$ as continuous function of comoving time, we also
expect that this ratio is approximately given by $ r\simeq
-{\lambda_d\over \lambda_m}$ in the vicinity of $t_0$ where
$w_d(t_0)=-1$ occurs.

As a result, near $w_d=-1$, the value of $r(t_0)$ is specified by
the constant parameters of the interaction. If the present value
of $w_d$ is believed to be $w_d\approx -1$ \cite{mone}, based on
astrophysical data \cite{coin}, we can get $-{\lambda_d\over
\lambda_m} \simeq 3/7$. In this method we cannot assess
$\lambda_d$ and $\lambda_m$ separately. Note that for interactions
which do not satisfy (\ref{26}) (e.g. models with ${\lambda_d\over
\lambda_m}>0$), $w_d=-1$ is not accessible.

To investigate the condition (\ref{22}), we note that at $w_d=-1$
\begin{equation}\label{27}
\dot{Q}(t_0)=8\pi \lambda_d(1+r)\rho_d^2.
\end{equation}
Hence the inequality (\ref{22}) becomes
\begin{equation}\label{28}
32\pi \lambda_d(1+r)\rho_d^2(t_0)\leq (V'(\phi))^2.
\end{equation}
Note that the above constraint depends on the form of the
potential of the quintessence field. We shall discuss this issue
shortly after some remarks.

The conditions like (\ref{16}) and (\ref{22}), are only necessary
conditions to reach $w_d=-1$. Indeed we have used the fact that
{\it{if}} in an interacting quintessence model $w_d=-1$ is
achieved, {\it{then}} equations like \ref{16} and \ref{22} must be
hold. But we did not prove that $w_d=-1$ is allowed in the model.
In fact describing the exact form of $w_d$ (to see whether
$w_d=-1$ is accessible) requires solving the equation (\ref{6})
with one of the Friedmann equations in (\ref{4}), which in the
presence of interaction (even in its absence), as we mentioned in
the introduction, is not straightforward for a general potential.
Despite this, to study more about the behavior of the system near
$w_d=-1$, we can restrict ourselves to the neighborhood of $t=t_0$
where the Hubble parameter is presumed to be differentiable and
consider the interaction (\ref{25}). For a differentiable (at
least in an open interval containing $t=t_0$) Hubble parameter, in
the vicinity of $t_0$ (an open set containing $t_0$) following
\cite{sad2} we use the following Taylor expansion
\begin{equation}\label{29}
H=h_0+h_1(t-t_0)^{\beta}+\mathcal{O}(t-t_0)^{\beta+1},\,\,\,
\beta\geq 1.
\end{equation}
In this open set we consider the following expressions (Taylor
expansion):
\begin{eqnarray}\label{30}
w_d&=&-1+w_0(t-t_0)^\alpha+\mathcal{O}(t-t_0)^{\alpha+1}\nonumber \\
r&=&r_0+r_1(t-t_0)^\gamma+\mathcal{O}(t-t_0)^{\gamma+1}.
\end{eqnarray}
The equation of state parameter of the system is related to $w_d$
through the relation $w=w_d\Omega_d$. We have also
$w=-1-{2\dot{H}\over 3H^2}$, therefore near $w_d=-1$,
$w=-\Omega_d$ and $\dot{H}\neq 0$, and $\beta$ in (\ref{29})
begins with $\beta=1$. We write the equation (\ref{7}) as
\begin{equation}\label{31}
\dot{r}=3rH\Big[w_d+{1\over 3}\left( {r+1\over
r}\right)(\lambda_d+r\lambda_m) \Big]
\end{equation}
Putting (\ref{29}) and (\ref{30}) into (\ref{31}) gives:
$\gamma=1$ and $r_1=-3r_0h_0$. To elucidate our results we must
specify the potential. Here we consider the quadratic and the
exponential potentials. For the quadratic potential
\begin{equation}\label{32}
V(\phi)={1\over 2}m^2\phi^2,
\end{equation}
by substituting (\ref{29}) and (\ref{30}) in (\ref{14}) we arrive
at
\begin{equation}\label{33}
\pm
2m\sqrt{(1-w_d^2)}=\dot{w_d}+3H(1-w_d^2)+H(\lambda_d+\lambda_mr)(1-w_d).
\end{equation}
Comparing the coefficients of the expressions with the same power
of $t$ in both sides of (\ref{33}) forces us to take $\alpha=2$
and $r_0=-{\lambda_d\over \lambda_m}$, in accordance with
(\ref{16}). We also obtain the equation
\begin{eqnarray}\label{34}
\pm 2m\sqrt{2w_0}&=&2w_0+2\lambda_mr_1h_0\nonumber \\
&=&2w_0-6\lambda_dh_0^2
\end{eqnarray}
in agrement with the previous result (\ref{21}). The above
equation has real roots provided that
\begin{equation}\label{35}
8\pi \lambda_d(1-{\lambda_d\over \lambda_m})\phi^2(t_0)\leq 1,
\end{equation}
which is the same as (\ref{28}). This poses a condition on the
value of the quintessence field at $w_d=-1$.  In terms of total
energy density this inequality may be written as
\begin{equation}\label{36}
16\pi \lambda_d \rho(t_0)\leq m^2.
\end{equation}

For the exponential potential
\begin{equation}\label{37}
V=v_0\exp(\lambda \phi),\,\, v_0>0,
\end{equation}
(\ref{14}) reduces to
\begin{equation}\label{38}
\mp
\lambda(1-w_d)\sqrt{\rho_d(1+w_d)}=\dot{w_d}+3H(1-w_d^2)+H(\lambda_d+\lambda_mr)(1-w_d).
\end{equation}
Again, by substituting (\ref{29})  and (\ref{30}) in (\ref{38}),
and by comparing the coefficients of the same power of $t$ in both
sides we arrive at: $\alpha=2$, $r_0=-{\lambda_d\over \lambda_m}$
and
\begin{equation}\label{39}
\mp \lambda \sqrt{\rho_{d}(t_0)w_0}=w_0+3\lambda_dh_0^2,
\end{equation}
in agrement with (\ref{16}) and (\ref{21}). The necessary
condition to have real roots for (\ref{39}) is then
\begin{equation}\label{40}
32\pi \lambda_d(1+r_0)<\lambda^2
\end{equation}
By taking $r_0=-{\lambda_d\over \lambda_m}\approx 3/7 $ as the
present estimated value, we obtain
\begin{equation}\label{41}
{320\pi\over 7}\lambda_d\leq \lambda^2,
\end{equation}
which shows that access to $w_d=-1$ is not feasible for models
with arbitrary $\lambda$ and $\lambda_d$.
\section{Conclusion}
In this paper we considered a spatially flat FRW universe composed
of dark matter and dark energy components. The dark energy was
assumed to be a quintessence scalar field interacting with dark
matter (see (\ref{2}) and (\ref{5})). A general expression for
time derivative of EoS parameter of dark energy was derived (see
(\ref{14})), upon which we discussed some necessary conditions and
relation between the interaction term and the potential of the
quintessence to reach at $w_d=-1$ (see (\ref{16}) and (\ref{22})).
We also examine our results by approximation method based on
series expansion near the time when $w_d=-1$.

\end{document}